\documentclass{ws-procs9x6-cpt16}
\usepackage{amsmath}
\begin{document}

\newcommand{\refeq}[1]{(\ref{#1})}
\def\etal {{\it et al.}}

\title{CPT-Violating, Massive Photons and Cherenkov Radiation}

\author{Don Colladay,$^1$}

\address{$^1$New College of Florida,\\
Satrasota, FL 34234, USA}

\begin{abstract}
CPT-Violating photons are well-known to have problems with energy positivity in certain cases and therefore have not been convincingly quantized to date.  We find that by adding a small mass term, consistent with experimental bounds, the theory can be regulated and allows for a consistent covariant quantization procedure.  This new framework is applied to a consistent quantum calculation of vacuum Cherenkov radiation rates.  These rates turn out to be largely independent of the mass of the photon regulator used.  In the physical regime, accessible by ultra high energy cosmic rays, the behavior of the rate is proportional to the square of the CPT-violating parameter and is not realistically observable.
\end{abstract}

\bodymatter

\section{Overview}
Vacuum Cherenkov Radiation is a generic feature of Lorentz-violating dynamics as there is often the possibility of particles obtaining speeds higher than the phase velocity of light in vacuum.
One may generally adopt two viewpoints regarding this effect.

\begin{enumerate}
    \item Cherenkov radiation is an instability of the theory due to the existence of negative-energy states in experimentally accessible frames (concordant frames \cite{ralph}) and is an indication that higher-order operators must become relevant to protect against these instabilities.  
    \item The theory is correct as it stands and the Cherenkov radiation is real and will happen in nature.
    Using this interpretation, the computation of rates can
    be used to place stringent bounds on certain Lorentz-violating parameters.
\end{enumerate}
In the first case, the radiation cannot be considered an actual observable effect, but it is an indication that physics must be modified at the appropriate energy scale likely leading to other observable effects.
In the second case, the absence of Cherenkov radiation from Ultra High Energy Cosmic Rays can place stringent bounds on certain parameters involving Lorentz violation.\cite{kostasson,klinkshreck,anselmi,jorge}

\section{CPT-violating, Massive Photons and QFT}

The specific goal of this talk is to present an overview of the Cherenkov effect when the time-like CPT-violation parameter $k_{AF}^\mu$ for which a term
\begin{equation}
{\cal{L}} \supset \frac 1 2 k_{AF}^\mu \epsilon_{\mu\nu\alpha\beta} A^\nu F^{\alpha\beta},
\end{equation}
is present in the photon sector.\cite{cfj, kostelecky-colladay}
Note that spacelike $k_{AF}^\mu$ does not exhibit similar problems and has 
been studied in detail elsewhere.\cite{lehnert-potting}
If this is the only term added to the conventional photon lagrangian, quantization is known to be problematic for at least two reasons.

\begin{enumerate}
\item There is a gap at low-energies in the energy-momentum relation indicating that there is no observer-Lorentz invariant way to separate particle and anti-particle states.
\item At certain momenta values the polarization vectors do not form a complete basis as required by the commutation relations of the fields. 
\end{enumerate}
Both of these issues can be remedied by inclusion of a mass term for the photon.  
In general, this mass only needs to dominate the $k_{AF}$ parameter so it can be chosen well-below current experimental limits and the theory can remain compatible with known physics.
An alternative approach is to work within the classical framework and compute the radiation in the absence of a mass term.  Curiously, this gives a zero result for the ratiation rate. \cite{altschul}

\section{Cherenkov Rate Calculation}
For simplicity, the results here are computed in a frame where $k^\mu_{AF} = (k^0,0,0,0)$.
There is a single diagram involving a charged fermion emitting a single photon with appropriate helicity such that $p^2 = m^2 - 2 k^0 |\vec p| < 0$, allowing for a nonzero Cherenkov radiation rate.  
The details of the calculation have recently been published. \cite{colpotmcd}

\begin{figure}
\begin{center}
\includegraphics[width=4in]{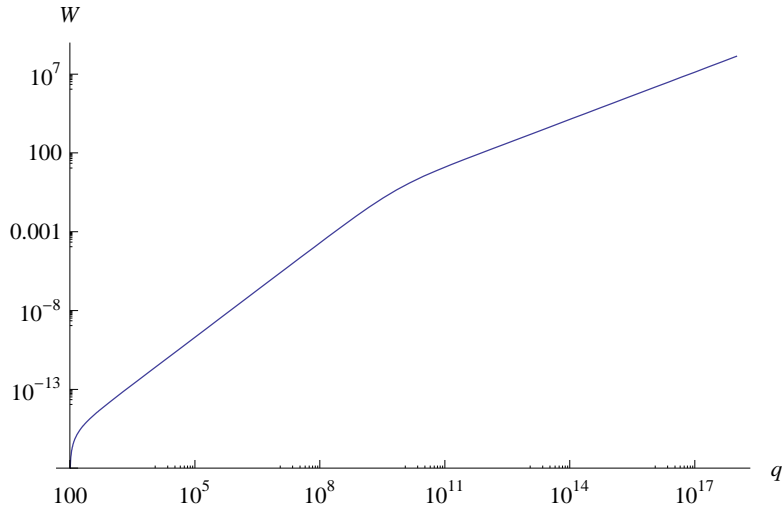}
\end{center}
\caption{Rate for Cherenkov radiation plotted for $m_\gamma/k^0 = 100$, in units of the mass.  Note that the threshold is very close to $q=100$ and the existence of two linear regions (on the log-log plot) in which the rate is proportional to $(kq)^2$ and $kq$ respectively.}
\label{aba:fig1}
\end{figure}

The computation can be performed exactly using mathematica, but the result is unwieldy and complicated.  Two interesting limits can be observed in the ranges of fermion momenta $\vec q$ where
\begin{align}
|\vec q| \gg \frac{m^2}{k_ {AF}^0}
&\qquad\Longrightarrow\qquad
W \sim {2 \over 3} \alpha \left( k_{AF}^0 \right) |\vec q|,
\nonumber\\
\frac{m\,m_\gamma}{k_ {AF}^0} \ll |\vec q| \ll \frac{m^2}{k_ {AF}^0}
&\qquad\Longrightarrow\qquad
W \sim \alpha \left({k_ {AF}^0 \over m}\right)^2|\vec q|^2,
\end{align}
where $m$ is the mass of a singly charged fermion and $\alpha$ is the fine structure constant.
These two regions are easily identified as the flat regions in Figure 1.

Making recourse to experimental values, $k_{AF} \sim 10^{-42}$GeV is bounded by cosmological birefringence measurements to be extremely small.\cite{datatables}
Taking the mass of the photon as two-orders of magnitude larger than this value is well-within the experimental bounds of $m_\gamma < 10^{-27}$GeV quoted by the particle data group.\cite{pdg}  Moreover, a mass of this scale is even much smaller than the more speculative bounds based on galactic magnetic fields on the order of $m_\gamma < 10^{-36}$GeV.\cite{photbounds}
Using the above values, the region relevant for experiment clearly lies where the rate is proportional to the square of the CPT-violating coefficient.  This indicates that a typical Ultra High Energy Cosmic Ray would take about 40 times the age of the universe to radiate a significant percentage of its total initial energy making the effect irrelevant for experiment.  This is interesting as the threshold can in fact be quite low (depending on $m_\gamma/k_0$, but the rate is so much suppressed as to make the effect of Cherenkov radiation irrelevant.)

\section{Conclusion}

CPT-violating electrodynamics in the minimal Standard Model Extension has traditionally had serious consistency problems as a quantum theory when the photon parameter $k_{AF}$ is time-like.  
Introduction of a small nonzero mass term for the photon can provide a remedy for these problems allowing for the calculation of the rate of Cherenkov radiation.  
While the threshold for emission depends on the size
of the photon mass relative to the CPT-violating parameter,
the actual rate ends up being largely independent of this mass yielding a "regulated" result for Cherenkov radiation. In the physical regime, the rate goes as the square of the CPT-violating coefficient yielding an un-observably slow rate of emission.  
The net result of this analysis of CPT-violation in the photon sector is that the existence of negative energy states in non-concordant frames may in fact not be as problematic as previously thought.  Instead of introducing additional higher-order operators to cure the instability, it can be possible that the rate of radiation emission is so slow as to be un-obervable, rendering the instabilities harmless from an experimental viewpoint.

\section*{Acknowledgments}
The author would like to thank New College of Florida's summer faculty development program for support used to travel to the conference.

\end{document}